\documentclass[conference]{IEEEtran}
\usepackage{cite}
\usepackage{amsmath,amssymb,amsfonts}
\usepackage{graphicx}
\usepackage{textcomp}
\usepackage{xcolor}
\usepackage{float}
\usepackage{amsthm}
\usepackage{algorithm}
\usepackage{algpseudocode}
\usepackage{bm}

\def\Htran{\mbox{\tiny $\mathrm{H}$}}
\def\Ttran{\mbox{\tiny $\mathrm{T}$}}

\newtheorem{remark}{Remark}

\IEEEoverridecommandlockouts 

\begin{document}

\title{Hardware Distortion Aware Precoding for ISAC Systems}

\author{\IEEEauthorblockN{Murat Babek Salman and Emil Bj{\"o}rnson}

\IEEEauthorblockA{KTH Royal Institute of Technology, Stockholm, Sweden\\
Email: \{mbsalman, emilbjo\}@kth.se}
\and
\IEEEauthorblockN{ Özlem Tuğfe Demir}
\IEEEauthorblockA{Bilkent University, Ankara, Türkiye\\
Email: ozlemtugfedemir@bilkent.edu.tr}%
\thanks{This work was supported by the Swedish Innovation Agency (Vinnova) through the SweWIN center (2023-00572). \"O. T. Demir was supported by 2232-B International Fellowship for Early Stage Researchers Programme funded by the Scientific and Technological Research Council of Turkiye.}}

\maketitle

\begin{abstract}
The impact of hardware impairments on the spectral efficiency of communication systems is well studied, but their effect on sensing performance remains unexplored. In this paper, we analyze the influence of hardware impairments on integrated sensing and communication (ISAC) systems in cluttered environments. We derive the sensing signal-to-clutter-plus-noise ratio (SCNR) and show that hardware distortions significantly degrade sensing performance by enhancing clutter-induced noise, which masks target echoes. The isotropic nature of transmit distortion due to multiple stream transmission further complicates clutter suppression. To address this, we propose a distortion- and clutter-aware precoding strategy that minimizes the deviation from the communication-optimized precoder while improving sensing robustness. We also propose an alternative power allocation-based approach that reduces computational complexity. Numerical results confirm the effectiveness of the proposed approaches in overcoming hardware- and clutter-induced limitations, demonstrating significant performance gains over distortion-unaware designs.
\end{abstract}


\section{Introduction}

Within the progress of sixth-generation (6G) communication technologies, integrated sensing and communication (ISAC) has emerged as a key feature, addressing demands for enhanced environmental awareness, security, and localization capabilities \cite{10217169}. By merging communication and radar functionalities over shared hardware and spectrum, ISAC offers improved efficiency and performance. MIMO (multiple-input multiple-output) technology further strengthens ISAC systems by enabling spatial diversity for reliable joint operations \cite{10124714}.

ISAC systems must balance high spectral efficiency (SE) for data transmission with reliable environmental sensing accuracy \cite{8828030}. Among various approaches, the joint design method has received increasing attention as it optimizes both sensing and communication performance \cite{8288677}. However, as future base stations (BSs) become equipped with larger antenna arrays (e.g., in the 6G upper mid-band), hardware complexity increases. A practical approach to enabling large-scale MIMO is to use cost-efficient, low-power components, which inevitably introduce hardware impairments \cite{6891254}. These impairments degrade both communication and sensing performance, and clutter-rich environments further amplify their impact by diminishing the effective signal-to-clutter-plus-noise ratio (SCNR).

Several previous studies have investigated hardware constraints in communication and sensing systems independently. The impact of 1-bit digital-to-analog converters (DACs) on communication is addressed in \cite{7967843}, while \cite{9780031} analyzes the degradation of radar performance due to coarse quantization. In ISAC systems, \cite{10794378} analyzes power amplifier nonlinearities on the sensing signal-to-noise ratio (SNR), and \cite{9399801} proposes nonlinear precoding under quantization constraints, but with high computational complexity.

Linear precoding techniques have been proposed to realize desired beampatterns for ISAC \cite{9124713,10086626}, although maintaining the intended beampatterns remains challenging due to isotropic hardware distortions. Other methods focus on minimizing the Cramér-Rao bound or maximizing sensing SNR \cite{9652071,9968163}. The work \cite{10571197} considers transmit distortion noise to maximize sensing SNR with communication constraints.

Despite these advancements, none of the aforementioned works explicitly consider receiver hardware impairments or account for clutter an interference source. These omissions are critical, as strong clutter returns can saturate the receiver front-end, amplifying hardware distortion effects and significantly degrading sensing performance.

In this work, a closed-form expression for sensing SCNR is derived, incorporating both transmit and receiver hardware distortions in cluttered environments.
A novel precoding framework is proposed to minimize deviation from the ideal communication-oriented precoder while ensuring sensing performance. We demonstrate through extensive simulations that distortion-aware design is essential for reliable ISAC operation under practical hardware constraints. 
    \vspace{-1mm}

\section{System Model}
We consider an ISAC system in which a BS, equipped with $M$ antennas for both transmission and reception, serves $K$ single-antenna downlink user equipments (UEs) and performs monostatic target detection.
\subsection{Transmitted Signal Model}

The $M$-dimensional transmitted signal from the antenna array at the BS at the $n^{\rm th}$ sample in a coherence block can be expressed as 
\begin{equation}
    \mathbf{x}[n]= \sqrt{\rho^{\rm tot}} \mathbf{W} \sqrt{\boldsymbol{\rho}} \mathbf{s}[n] + \sqrt{\rho^{\rm tot}\rho_0} \mathbf{w}_{0} s_0[n] + \boldsymbol{\eta}_{\text{\tiny t}}[n],
\end{equation}
where $\mathbf{s}[n] = [s_1[n], \ldots, s_K[n]]^{\Ttran}$ is the communication symbol vector with $\mathbb{E}[\mathbf{s}[n]\mathbf{s}^{\Htran}[n]] = \mathbf{I}_K$, $\rho^{\rm tot}$ is the total transmit power, $\boldsymbol{\rho} = \operatorname{diag}(\rho_1,\ldots,\rho_K)$ contains the UE power coefficients, $\mathbf{W} \in \mathbb{C}^{M\times K}$ is the precoding matrix for the communication signal, $s_0[n]$ is the dedicated radar waveform with $\mathbb{E}[|s_0[n]|^2] = 1$, $\rho_0$ is the power coefficient for target detection, and $\mathbf{w}_{0} \in \mathbb{C}^M$ is the beamforming vector for the sensing task. The power coefficients satisfy $\sum_{k=0}^K \rho_k = 1$.

Due to imperfections in the hardware, the transmitted signal is impaired by hardware distortion, represented by $\boldsymbol{\eta}_{\text{\tiny t}}[n] \sim \mathcal{CN}(\mathbf{0}, \mathbf{R}_{\text{\tiny t}})$, which is uncorrelated with the intended signal. In this work, we adopt the uncorrelated additive distortion model, where its second-order statistics can be modeled as $\mathbf{R}_{\text{\tiny t}} = \kappa_{\text{t}} \frac{\rho^{\rm tot}}{M} \mathbf{I}_{M}$, where $\kappa_{\text{t}} \geq 0$ is the hardware distortion factor \cite{6891254}. The dependence on $\rho^{\rm tot}$ makes the hardware distortion much different than the receiver noise. It should be noted that the uncorrelated distortion model becomes practical if the number of transmitted streams increases, since the distortion is radiated in approximately $(K+1)^2$ directions \cite{6891254}.

    \vspace{-1mm}

\subsection{Received Communication Signal}

The received signal at the $k^{\rm th}$ communication UE is
    \vspace{-1mm}
\begin{equation}
    y_k^{\text{\tiny UE}}[n] = \mathbf{h}_k^{\Htran} \mathbf{x}[n] + \nu_k[n], \label{eq:CommRec}
\end{equation}
where $\mathbf{h}_k \in \mathbb{C}^{M \times 1}$ is the communication channel vector, modeled as correlated Rayleigh fading with distribution $\mathbf{h}_k \sim \mathcal{CN}(\mathbf{0}, \mathbf{R}_k)$, and $\nu_k[n]$ is additive white Gaussian noise (AWGN) with variance $\sigma_{\nu}^2$. The received signal in \eqref{eq:CommRec} can be expressed as
\begin{align}
    y_k^{\text{\tiny UE}}[n] =& \underbrace{\sqrt{\rho^{\rm tot}}\sqrt{\rho_k}\mathbf{h}_k^{\Htran}\mathbf{w}_k s_k[n]}_{\text{Intended signal}} + \underbrace{\sqrt{\rho^{\rm tot}}\sum_{i \neq k}^K \sqrt{\rho_i}\mathbf{h}_k^{\Htran}\mathbf{w}_i s_i[n]}_{\text{Multi-user interference}} \nonumber \\
    &+ \underbrace{\sqrt{\rho^{\rm tot}}\sqrt{\rho_0} \mathbf{h}_k^{\Htran} \mathbf{w}_{0} s_0[n]}_{\text{Sensing interference}} + \underbrace{\mathbf{h}_k^{\Htran} \boldsymbol{\eta}_{\text{\tiny t}}[n]}_{\text{Distortion noise}} + \nu_k[n].
\end{align}

The instantaneous communication signal-to-interference-plus-noise ratio (SINR) for the $k^{\rm th}$ communication UE for a single channel realization can be expressed as
\begin{equation}
    \Upsilon_k = \frac{\rho_k |\mathbf{h}_k^{\Htran} \mathbf{w}_k|^2}{\sum_{i \neq k}^K \rho_i |\mathbf{h}_k^{\Htran} \mathbf{w}_i|^2 + \rho_0 |\mathbf{h}_k^{\Htran} \mathbf{w}_{0}|^2 + \frac{\kappa_{\text{t}}}{M} \|\mathbf{h}_k\|^2 + \frac{\sigma_{\nu}^2}{\rho^{\rm tot}}}. \label{eq:CommSNR}
\end{equation}

It is important to note that the effective distortion power is primarily determined by the total transmit power, regardless of how it is allocated between communication and sensing signals. As a result, increasing the sensing power reduces the power available for communication, while still contributing to the same overall distortion level. This trade-off leads to a degradation in communication performance. Therefore, managing the effective distortion power by carefully balancing the power split between sensing and communication is crucial to maintaining reliable communication quality in hardware-impaired systems.

\subsection{Received Echo Signal}

The BS processes the received echo signal to detect a possible target. The received echo signal can be expressed as
\begin{equation}
\begin{split}
    \mathbf{y}^{\text{\tiny BS}}[n] =& \alpha_0 \mathbf{a}(\theta_0)\mathbf{a}^{\Ttran}(\theta_0) \mathbf{x}[n] \\
    &+ \sum_{q=1}^Q \alpha_q \mathbf{a}(\theta_q)\mathbf{a}^{\Ttran}(\theta_q) \mathbf{x}[n-\tau_q] + \boldsymbol{\eta}_{\text{\tiny r}}[n] + \boldsymbol{\mu}[n], \label{eq:RXsensing}
\end{split}
\end{equation}
where $\mathbf{a}(\theta_0) = [1, e^{j\pi\sin\theta_{0}}, \ldots, e^{j\pi(M-1)\sin\theta_{0}}]^{\Ttran}$ is the array response vector toward the target at angle $\theta_0$, $\mathbf{a}(\theta_q)$ is the array response vector for the $q^{\rm th}$ clutter at angle $\theta_q$, and $\tau_q$ is the relative delay of the $q^{\rm th}$ clutter echo with respect to the target.

The coefficient $\alpha_0$ captures the combined effects of the target radar cross-section (RCS) and the two-way propagation pathloss between the BS and the target. Similarly, $\alpha_q$ denotes the complex gain associated with the $q^{\rm th}$ clutter source, which accounts for both the reflection characteristics and the propagation effects of the respective clutter path. The average power of each component is characterized by its second-order moment, given by $\mathbb{E}[|\alpha_q|^2] = \bar{\alpha}_q^2$, for $q = 0, \dotsc, Q$.

The term $\boldsymbol{\eta}_{\text{\tiny r}}[n] \sim \mathcal{CN}(\mathbf{0}, \mathbf{R}_{\text{\tiny r}})$ represents hardware distortion caused by the imperfect receiver hardware. The autocorrelation matrix of the receiver distortion noise can be represented as $\mathbf{R}_{\text{\tiny r}} = \kappa_{\text{r}} \operatorname{diag}(\mathbb{E}[\mathbf{z}^{\text{\tiny BS}}[n] (\mathbf{z}^{\text{\tiny BS}}[n])^{\Htran}])$, where  $\mathbf{z}^{\text{\tiny BS}}[n] = \alpha_0 \mathbf{a}(\theta_0)\mathbf{a}^{\Ttran}(\theta_0) \mathbf{x}[n] + \sum_{q=1}^Q \alpha_q \mathbf{a}(\theta_q)\mathbf{a}^{\Ttran}(\theta_q) \mathbf{x}[n-\tau_q]$, which simplifies to
\begin{equation}
    \mathbf{R}_{\text{\tiny r}} = \kappa_{\text{r}} \sum_{q=0}^{Q} \bar{\alpha}_q^2 \rho^{\rm tot} (MP(\theta_q) + \kappa_{\text{t}}) \mathbf{I}_M,
\end{equation}
where $P(\theta_q)$ represents the beampattern of the transmitted waveform, defined as
\begin{equation}
    P(\theta) \triangleq \frac{1}{M} \mathbf{a}^{\Ttran}(\theta) \left(\tilde{\mathbf{W}}\tilde{\boldsymbol{\rho}}\tilde{\boldsymbol{\rho}}^{\Htran}\tilde{\mathbf{W}}^{\Htran}\right) \mathbf{a}^{*}(\theta),
\end{equation}
with $\tilde{\mathbf{W}} = [\mathbf{W}, \mathbf{w}_0]$ and $\tilde{\boldsymbol{\rho}} = \operatorname{diag}(\sqrt{\rho_1}, \ldots, \sqrt{\rho_K}, \sqrt{\rho_0})$. The term $\boldsymbol{\mu}[n] \sim \mathcal{CN}(\mathbf{0}, \sigma_\mu^2 \mathbf{I}_M)$ represents thermal receiver noise.

\begin{remark}
The receiver distortion term $\boldsymbol{\eta}_{\text{\tiny r}}[n]$ depends on the total received power, which includes contributions from both the target and clutter echoes. In scenarios with strong clutter, the receiver distortion power can become significant, potentially dominating the noise floor and severely degrading sensing performance. This coupling between clutter power and receiver distortion is a key challenge in hardware-impaired ISAC systems.
\end{remark}

We will analyze how transceiver hardware impairments, particularly under strong clutter interference, impact the ability to detect targets in the considered system setup. A widely used performance metric in this context is the sensing SCNR, which is a meaningful parallel to classical communication metrics.

\section{Performance Analysis and Precoder Design} \label{sec:PerfAnPre}

In this section, we first present a detailed performance analysis where we derive a closed-form expression for the sensing SCNR, explicitly accounting for hardware impairments and clutter interference. Building on this analytical result, we formulate a novel precoding optimization problem that aims to ensure a desired sensing performance while considering communication performance.

\subsection{Sensing SCNR Derivation}

The sensing SCNR can be written in the general form shown in \eqref{eq:sensingSNR} at the top of the next page, where $\boldsymbol{\omega}$ is the receive combining vector.
\begin{figure*}[t]
\begin{equation}
    \Upsilon_0 = \frac{\rho^{\rm tot}\bar{\alpha}_0^2 \boldsymbol{\omega}^{\Htran} \left[\mathbf{a}(\theta_0)\mathbf{a}^{\Ttran}(\theta_0) \tilde{\mathbf{W}}\tilde{\boldsymbol{\rho}} \tilde{\boldsymbol{\rho}}^{\Htran} \tilde{\mathbf{W}}^{\Htran} \mathbf{a}^{*}(\theta_0) \mathbf{a}^{\Htran}(\theta_0)\right]\boldsymbol{\omega}}{\boldsymbol{\omega}^{\Htran} \sum_{q=1}^Q \bar{\alpha}_q^2 \mathbf{a}(\theta_q)\mathbf{a}^{\Ttran}(\theta_q) \left(\rho^{\rm tot}\tilde{\mathbf{W}}\tilde{\boldsymbol{\rho}} \tilde{\boldsymbol{\rho}}^{\Htran} \tilde{\mathbf{W}}^{\Htran} + \mathbf{R}_{\text{\tiny t}}\right) \mathbf{a}^{*}(\theta_q)\mathbf{a}^{\Htran}(\theta_q)\boldsymbol{\omega} + (\sigma_\mu^2 + \kappa_{\text{r}} \sum_{q=0}^{Q} \bar{\alpha}_q^2 \rho^{\rm tot} (MP(\theta_q) + \kappa_{\text{t}}))\boldsymbol{\omega}^{\Htran}\boldsymbol{\omega}} \label{eq:sensingSNR}
\end{equation}
\hrulefill
\end{figure*}
There are two alternative approaches for designing the combiner. The first alternative is the clutter-aware combiner design, which can be expressed as
\begin{equation}
    \boldsymbol{\omega} = \left(\sum_{q=1}^Q \sigma^2_q \mathbf{a}(\theta_q) \mathbf{a}^{\Htran}(\theta_q) + \bar{\sigma}_\mu^2 \mathbf{I}_M\right)^{-1} \mathbf{a}(\theta_0), 
    \label{eq:statComb}
\end{equation}
where $\sigma^2_q \triangleq \bar{\alpha}_q^2 \rho^{\rm tot}(MP(\theta_q) + \kappa_{\text{t}})$ and $\bar{\sigma}_\mu^2 \triangleq \sigma_{\mu}^2 + \kappa_{\text{r}} \sum_{q=0}^Q \rho^{\rm tot}\bar{\alpha}_q^2(MP(\theta_q)+\kappa_{\text{t}})$. 
The combiner in \eqref{eq:statComb} attempts to suppress interference across the clutter regions. The second alternative is to apply the spatial matched filter along the target direction, i.e., $\boldsymbol{\omega} = \mathbf{a}(\theta_0)$. Note that the matched filter combiner maximizes the SNR under AWGN interference; however, under spatially correlated clutter interference, its performance significantly degrades.

The sensing SCNR for the clutter-aware combiner can be written as
\begin{equation}
\begin{split}
    \Upsilon_0 &= \sigma_0^2 \mathbf{a}^{\Htran}(\theta_0)\left(\sum_{q=1}^Q \sigma^2_q \mathbf{a}(\theta_q) \mathbf{a}^{\Htran}(\theta_q) + \bar{\sigma}_\mu^2 \mathbf{I}_M\right)^{-1} \mathbf{a}(\theta_0) \\
    &= \sigma_0^2 \mathbf{a}^{\Htran}(\theta_0)\left(\mathbf{A} \boldsymbol{\Sigma} \mathbf{A}^{\Htran} + \bar{\sigma}_\mu^2 \mathbf{I}_M\right)^{-1} \mathbf{a}(\theta_0), \label{eq:MMSESNR}
\end{split}
\end{equation}
where $\sigma_0^2 \triangleq \bar{\alpha}_0^2 \rho^{\rm tot} MP(\theta_0)$, $\mathbf{A} \triangleq [\mathbf{a}(\theta_1), \ldots, \mathbf{a}(\theta_Q)]$, and $\boldsymbol{\Sigma} \triangleq \operatorname{diag}(\sigma_1^2, \ldots, \sigma_Q^2)$.
By using the matrix inversion lemma, we can re-express \eqref{eq:MMSESNR} as
\begin{equation}
    \Upsilon_0 = \frac{\sigma_0^2}{\bar{\sigma}_\mu^2} \mathbf{a}^{\Htran}(\theta_0) \left(\mathbf{I}_M - \mathbf{A} \left(\bar{\sigma}_\mu^2\boldsymbol{\Sigma}^{-1} + \mathbf{A}^{\Htran} \mathbf{A}\right)^{-1} \mathbf{A}^{\Htran}\right) \mathbf{a}(\theta_0).
\end{equation}
This expression can be simplified further by assuming that clutter sources are separated in the angular domain and the number of antennas is large, yielding $\mathbf{A}^{\Htran}\mathbf{A} \approx M\mathbf{I}_{Q}$, which leads to
\begin{equation}
    \Upsilon_0 \approx \frac{\sigma_0^2}{\bar{\sigma}_\mu^2} \mathbf{a}^{\Htran}(\theta_0) \left(\mathbf{I}_M - \mathbf{A} \bar{\boldsymbol{\Sigma}} \mathbf{A}^{\Htran}\right) \mathbf{a}(\theta_0),
\end{equation}
where the $q^{\rm th}$ diagonal element of $\bar{\boldsymbol{\Sigma}}$ is $[\bar{\boldsymbol{\Sigma}}]_q \triangleq \frac{\sigma_q^2}{M\sigma_q^2+\bar{\sigma}_\mu^2}$. The sensing SCNR can then be written as
\begin{equation}
    \Upsilon_0 = \frac{\sigma_0^2}{\bar{\sigma}_\mu^2} \left(M - \sum_{q=1}^Q \frac{\sigma_q^2}{M\sigma_q^2+\bar{\sigma}_\mu^2} |\mathbf{a}^{\Htran}(\theta_0) \mathbf{a}(\theta_q)|^2\right).
\end{equation}

The sensing SCNR admits further simplification under the assumption of large-scale antenna arrays and sufficient angular separation between the target and clutter sources. In such a regime, the cross-correlation terms $|\mathbf{a}^{\Htran}(\theta_0) \mathbf{a}(\theta_q)|^2$ become negligible compared to $M$, yielding
\begin{equation}
    \Upsilon_0 = \frac{M \bar{\alpha}_0^2 \rho^{\rm tot} MP(\theta_0)}{\sigma_{\mu}^2 + \kappa_{\text{r}} \sum_{q=0}^Q \rho^{\rm tot}\bar{\alpha}_q^2(MP(\theta_q)+\kappa_{\text{t}})}. \label{eq:FinalSCNR}
\end{equation}

The final expression in \eqref{eq:FinalSCNR} highlights the dependency on the array size, beamforming gain, and total transmit power, which collectively enhance the received echo power from the desired target. In contrast, the denominator captures the impact of thermal noise, residual hardware impairments at both the transmitter and receiver, as well as interference contributions from all considered angular bins.

\subsection{Direct Precoder Design}

The proposed objective of this method is to minimize the deviation from the ideal communication-centric precoder while satisfying the sensing SCNR requirement. For this purpose, we consider the minimum mean-squared error (MMSE) precoding for communication, which can be expressed as
\begin{align}
    \bar{\mathbf{W}}^{\text{\tiny MMSE}} &= \left(\mathbf{H} \mathbf{H}^{\Htran} + \sigma_{\nu}^2 \mathbf{I}_M\right)^{-1} \mathbf{H}, \\
    \mathbf{w}^{\text{\tiny MMSE}}_k &=\sqrt{\rho_k}\frac{\bar{\mathbf{w}}^{\text{\tiny MMSE}}_k}{\sqrt{(\bar{\mathbf{w}}_k^{\text{\tiny MMSE}})^{\Htran} \bar{\mathbf{w}}^{\text{\tiny MMSE}}_k}},
\end{align}
where $\mathbf{H} \triangleq [\mathbf{h}_1, \ldots, \mathbf{h}_K]$, and the power allocation coefficients are set to $\rho_k = 1/K$ for $k=1,\ldots,K$ and $\rho_0=0$.

Then the following optimization problem can be formulated:
\begin{subequations}
\begin{align}
    \text{P1:} \quad &\underset{\mathbf{W}}{\text{minimize}} \quad \|\mathbf{W}^{\text{\tiny MMSE}} - \mathbf{W}\|_F \label{eq:P1-obj} \\
    &\text{subject to} \quad \Upsilon_0(\mathbf{W}) \geq \Gamma_{0}, \label{eq:P1-const1} \\
    &\qquad\qquad\quad \operatorname{trace}(\mathbf{W}\mathbf{W}^{\Htran}) \leq 1. \label{eq:P1-const2}
\end{align}
\end{subequations}

The non-convex constraint in \eqref{eq:P1-const1} has a fractional structure, which makes it infeasible to obtain a globally optimal solution in polynomial time. To obtain a solution to (P1), we will reformulate the constraint \eqref{eq:P1-const1} by rearranging the closed-form expression of the sensing SCNR constraint, isolating the precoding-dependent terms and bringing all terms to one side to express it as a standard quadratic inequality in $\{\mathbf{w}_k\}$ as
\begin{equation}
    \sigma_1^2 + \sum_{q=0}^Q d_q \sum_{k=1}^K |\mathbf{a}^{\Ttran}(\theta_q) \mathbf{w}_k|^2 - c_0\sum_{k=1}^K |\mathbf{a}^{\Ttran}(\theta_0) \mathbf{w}_k|^2 \leq 0, \label{eq:P1-newConst}
\end{equation}
where $c_0 \triangleq M \bar{\alpha}_0^2 \rho^{\rm tot}$, $d_q \triangleq \Gamma_0 \rho^{\rm tot}\kappa_{\text{r}} \bar{\alpha}_q^2$, and $\sigma_1^2 = \Gamma_0 \sigma_{\mu}^2 + \sum_{q=0}^Q \bar{\alpha}^2_q \Gamma_0\kappa_{\text{r}} \rho^{\rm tot}\kappa_{\text{t}}$.

Note that the updated constraint in \eqref{eq:P1-newConst} is the difference of two convex functions. It can be transformed into a convex optimization problem using the first-order Taylor approximation of the concave part around a given initial point. Consequently, the problem to be solved at the $i^{\rm th}$ iteration can be written as
\begin{subequations}
\begin{align}
    \text{P2}^{(i)}: &\underset{\mathbf{W}}{\text{minimize}} \quad \|\mathbf{W}^{\text{\tiny MMSE}} - \mathbf{W}\|_F \label{eq:P2-obj} \\
    &\text{s.t.} \quad \sigma_1^2 + \sum_{q=0}^Q d_q \sum_{k=1}^K |\mathbf{a}^{\Ttran}(\theta_q) \mathbf{w}_k|^2 + c_0\sum_{k=1}^K |\mathbf{a}^{\Ttran}(\theta_0) \mathbf{w}_k^{(i)}|^2 \nonumber \\
    &- 2c_0 \sum_{k=1}^K \operatorname{Re}\left\{\mathbf{a}^{\Ttran}(\theta_0) \mathbf{w}_k^{(i)}\mathbf{w}_k^{\Htran} \mathbf{a}^*(\theta_0)\right\} \leq 0, \label{eq:P2-const1} \\
    &\qquad \operatorname{trace}(\mathbf{W}\mathbf{W}^{\Htran}) \leq 1, \label{eq:P2-const2}
\end{align}
\end{subequations}
where $\mathbf{w}_k^{(i)}$ is the solution obtained from the previous iteration.

\subsection{Power Allocation-Based Precoder Design}

In scenarios where a dedicated sensing signal is needed, the problem can be recast as a power allocation problem. The precoding vector for each UE is fixed as
\begin{equation}
    \mathbf{w}_k = \frac{\bar{\mathbf{w}}_k^{\text{\tiny MMSE}}}{\|\bar{\mathbf{w}}_k^{\text{\tiny MMSE}}\|_2},
\end{equation}
and the sensing precoding employs a spatial matched filter projected onto the null space of clutter regions
\begin{equation}
    \bar{\mathbf{w}}_0 = \left(\mathbf{I}_M + \sum_{q=1}^Q \mathbf{a}^*(\theta_q)\mathbf{a}^{\Ttran}(\theta_q)\right)^{-1} \mathbf{a}^{*}(\theta_0)
\end{equation}
and we set $\mathbf{w}_0= \bar{\mathbf{w}}_0/\Vert\bar{\mathbf{w}}_0\Vert_2$. The sensing SCNR can be expressed as
\begin{equation}
    \Upsilon_0 = \frac{\mathbf{c}^{\Ttran} \bar{\boldsymbol{\rho}}}{\sigma_{2}^2 + \mathbf{d}^{\Ttran}\bar{\boldsymbol{\rho}}},
\end{equation}
where $\bar{\boldsymbol{\rho}} = [\rho_1, \ldots, \rho_K, \rho_0]^{\Ttran}$, $[\mathbf{c}]_k = M \bar{\alpha}_0^2 \rho^{\rm tot} |\mathbf{a}^{\Ttran}(\theta_0) \mathbf{w}_k|^2$, $[\mathbf{d}]_k = \kappa_{\text{r}}\rho^{\rm tot} \sum_{q=0}^Q \bar{\alpha}_q^2|\mathbf{a}^{\Ttran}(\theta_q) \mathbf{w}_k|^2$, and $\sigma_{2}^2 = \sigma_{\mu}^2 + \kappa_{\text{r}}\kappa_{\text{t}}\rho^{\rm tot}\sum_{q=0}^Q\bar{\alpha}_q^2$.

The optimization problem becomes
\begin{subequations}
\begin{align}
    \text{P3:} \quad &\underset{\boldsymbol{\rho}}{\text{minimize}} \quad \sum_{k=1}^K (\bar{\rho}_k - \rho_k)^2 \label{eq:P4-obj} \\
    &\text{subject to} \quad (\Gamma_0\mathbf{d} - \mathbf{c})^{\Ttran} \bar{\boldsymbol{\rho}} + \Gamma_0\sigma_{2}^2 \leq 0, \label{eq:P4-const1} \\
    &\qquad\qquad\quad \mathbf{1}^{\Ttran} \bar{\boldsymbol{\rho}} \leq 1, \label{eq:P4-const2}
\end{align}
\end{subequations}
which is a convex quadratic program (QP) with linear constraints that can be solved efficiently using general-purpose convex solvers. Here, $\bar{\rho}_k$ is the desired power allocation coefficient from communication-centric design.

\section{Performance Evaluations} \label{sec:PerfEval}

In this section, we evaluate the performance of the proposed methods under various system conditions. The communication center frequency is set to $f_c = 28$\,GHz, and $K = 8$ single-antenna UEs are uniformly distributed within a cell of radius 1\,km, with the angular spread of the UE channels set to $5^{\circ}$. The target is assumed to reside within a region located at a distance between 400 and 500\,m from the BS. Additionally, the environment contains $Q = 5$ clutter sources positioned near the BS, with distances uniformly distributed between 20 and 100\,m. The RCS power of each clutter source is randomly drawn from a uniform distribution in the range 10 to 20\,dB. The bandwidth of the transmitted signal is 50\,MHz, the noise variance is set to $-90$\,dBm, and the RCS of the target is $\sigma_{\rm RCS}^2 = 1$\,m$^2$. The two-way pathlosses for the target signal and clutter are determined via the radar equation as $\frac{c^2}{(4\pi)^3f_c^2 d_q^4}$, where $d_q$ is the distance.

\begin{figure}[t]
    \centering
    \includegraphics[width=0.95\columnwidth]{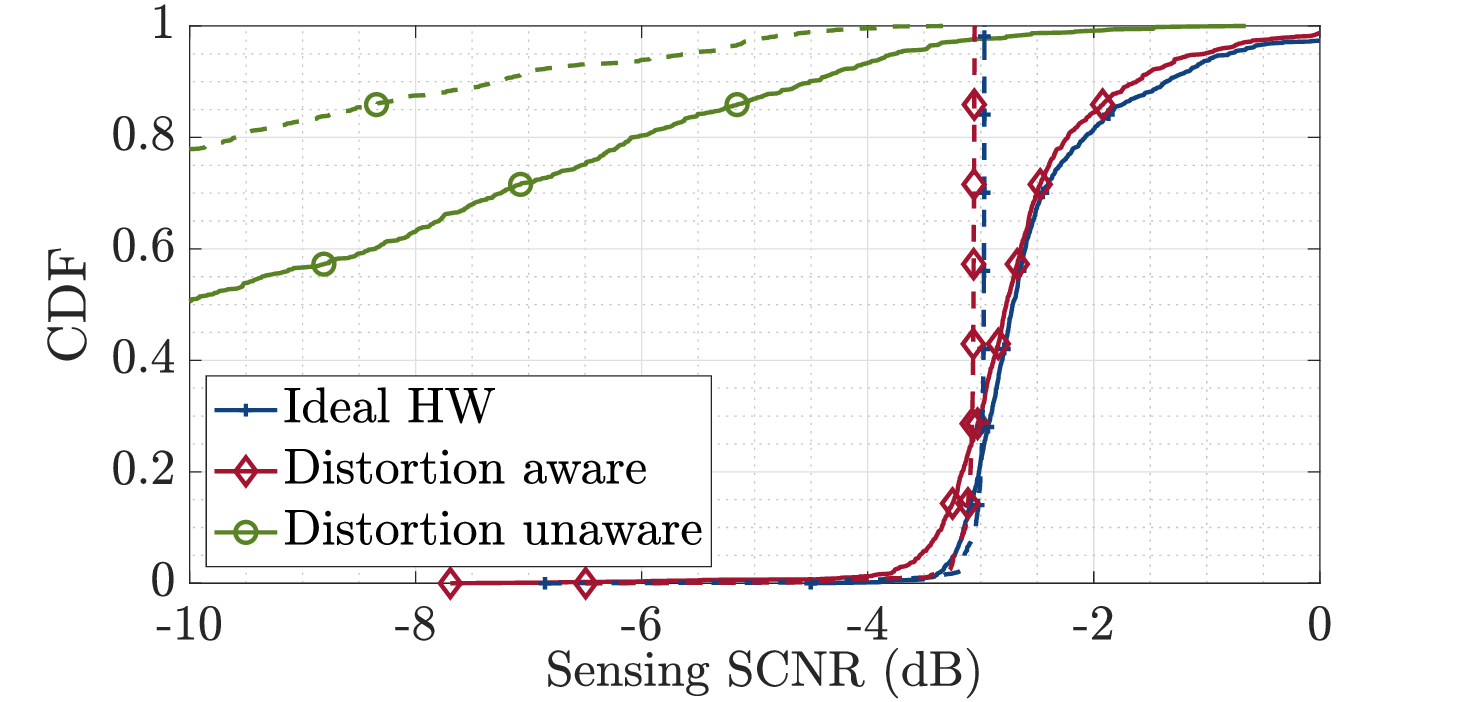}
    \caption{CDF of sensing SCNR (solid lines: precoding, dashed lines: power allocation).}
    \label{fig:CDFSCNRM100}
    \vspace{-4mm}
\end{figure}

The effectiveness of the proposed precoding strategy is validated by analyzing the cumulative distribution function (CDF) of the sensing SCNR for different UE and target locations in Fig.~\ref{fig:CDFSCNRM100}, with target SCNR set to $\Gamma_0 = 0.5$ for $\kappa_{\rm r}=\kappa_{\rm t}=0.01$. The proposed distortion-aware precoder successfully meets the SCNR requirement. In contrast, a distortion-unaware precoder, designed under the assumption $\kappa_{\text{r}} = \kappa_{\text{t}} = 0$, exhibits significant performance degradation. This highlights the importance of explicitly accounting for hardware-induced nonlinear distortion in the precoder design.

The distortion-aware power allocation can satisfy the sensing SCNR requirement very precisely. However, the proposed precoding method can achieve a higher sensing SCNR in scenarios where a communication UE is spatially close to the sensing target. In such cases, a significant portion of the transmit power is naturally directed toward the target without requiring an additional dedicated beam.

\begin{figure}[t]
    \centering
    \includegraphics[width=0.95\columnwidth]{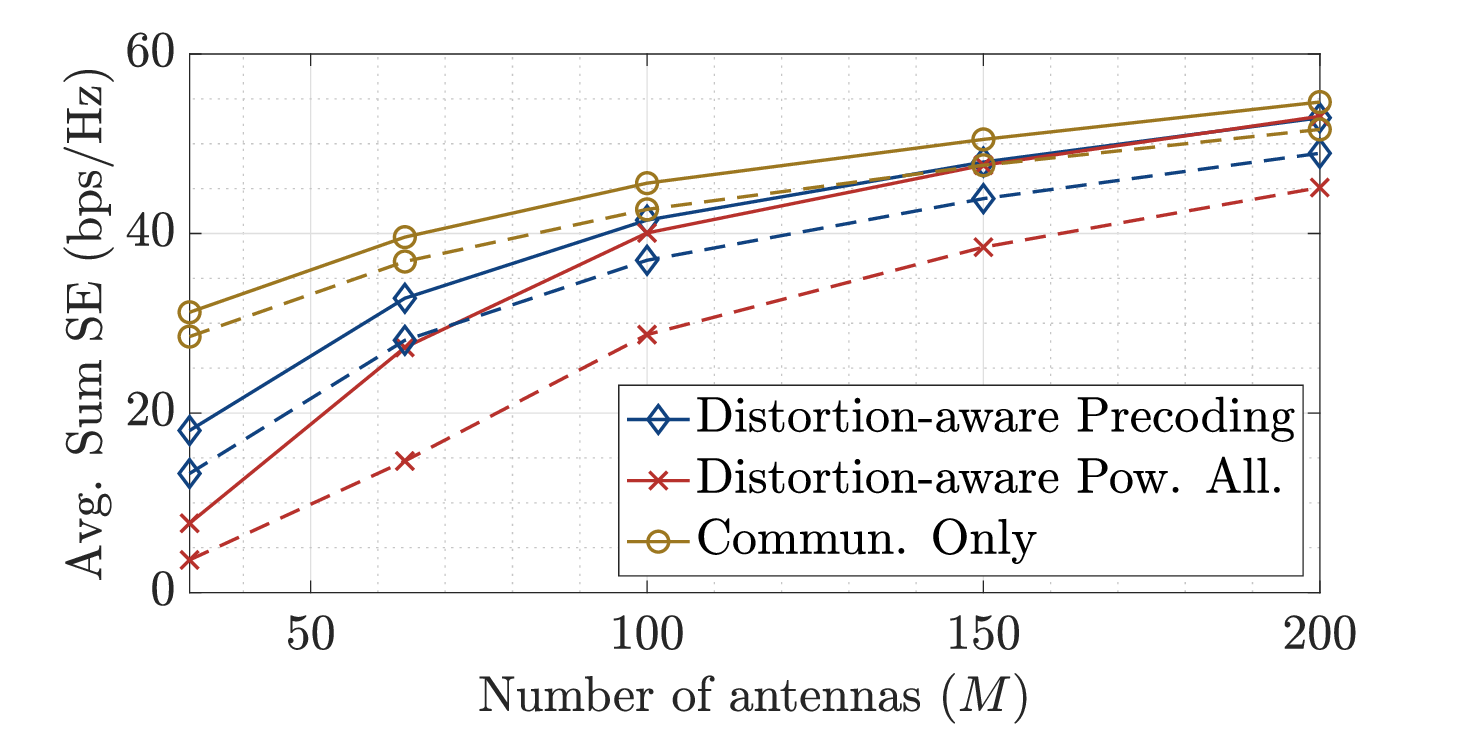}
    \caption{Average sum SE vs. number of antennas $M$ (solid lines: ideal hardware, dashed lines: imperfect hardware).}
    \label{fig:SEvsM}
        \vspace{-4mm}
\end{figure}

The effects of the number of antennas are examined in Fig.~\ref{fig:SEvsM}. The communication SE performance loss due to joint communication and sensing precoding tends to decrease as the number of antennas increases. Furthermore, hardware distortion affects the power allocation method more significantly compared to the precoding method as the number of antennas decreases. The performance degradation at lower antenna numbers arises because, to satisfy the sensing SCNR constraint, a larger portion of the transmit power must be directed toward the sensing target rather than the communication UEs.

As the number of antennas increases, the deviation from the ideal communication precoding becomes smaller, thanks to the enhanced array gain and improved capability to form sharp beams and nulls. This allows the system to better steer the beam toward the target and suppress clutter without severely compromising communication performance.

\begin{figure}[t]
    \centering
    \includegraphics[width=0.95\columnwidth]{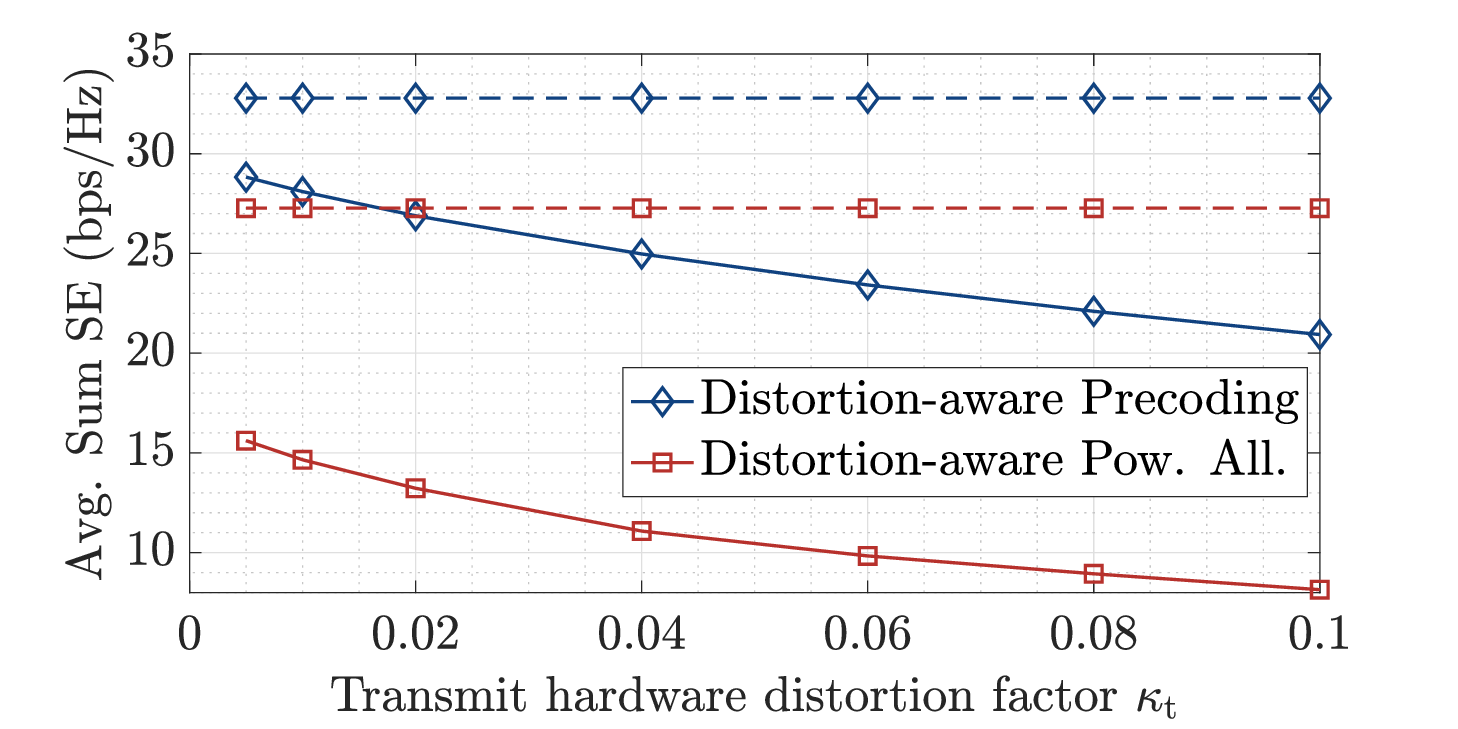}
    \caption{Average sum SE vs. transmit hardware distortion factor $\kappa_{\text{t}}$ (dashed lines: ideal hardware, solid lines: imperfect hardware).}
    \label{fig:SEvsKappat}

\end{figure}

The impact of transmit hardware distortion is investigated in Fig.~\ref{fig:SEvsKappat} for $M = 64$ transmit antennas with $\kappa_{\text{r}} = 0.01$. For lower distortion factors, the performance degradation is mild; however, as hardware distortion increases, the performance loss also increases. This is due to two reasons: first, the transmitted communication signal is distorted, which distorts the received UE signal; second, the beamformer is modified more to satisfy the sensing SCNR, which decreases the received desired communication signal power and increases interference.

\begin{figure}[t]
    \centering
    \includegraphics[width=0.95\columnwidth]{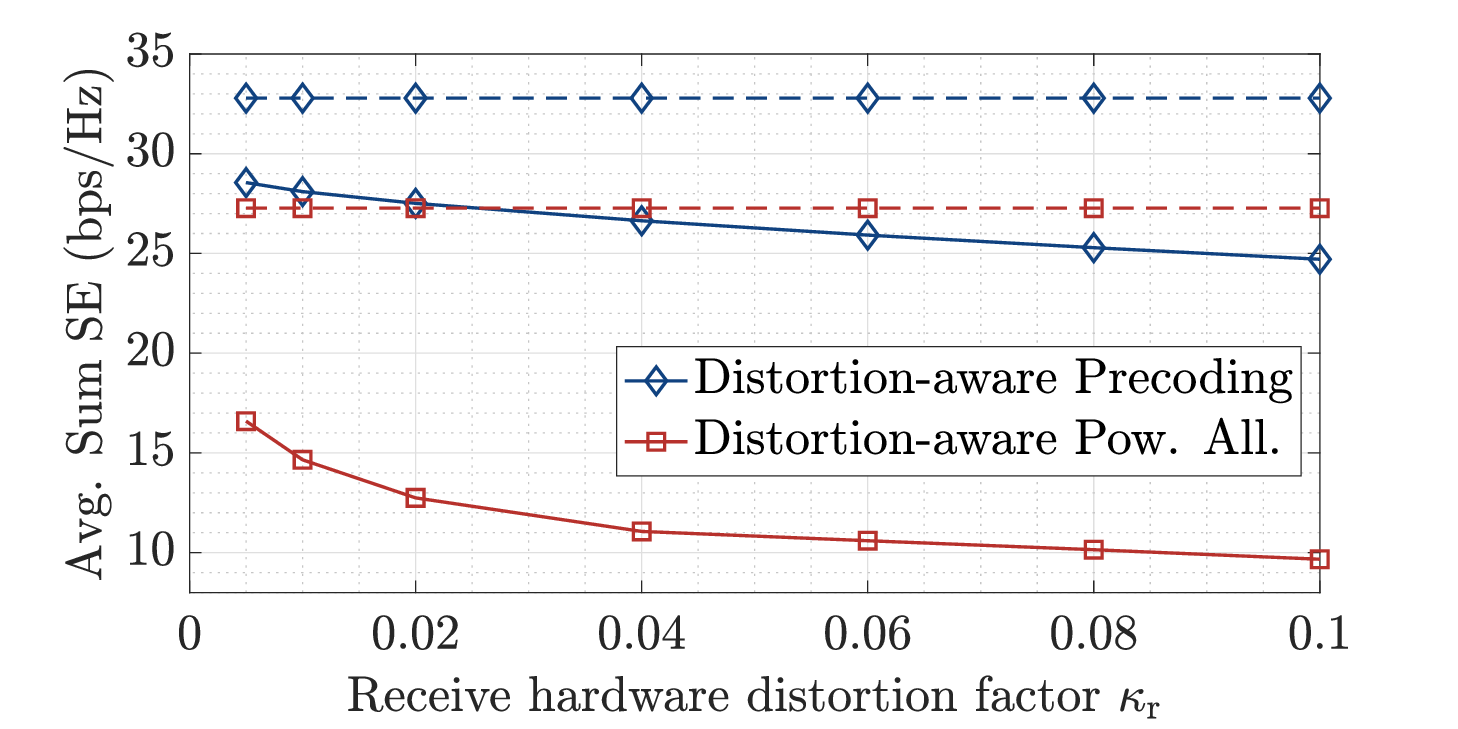}
    \caption{Average sum SE vs. receive hardware distortion factor $\kappa_{\text{r}}$ (dashed lines: ideal hardware, solid lines: imperfect hardware).}
    \label{fig:SEvsKappar}
        \vspace{-4mm}
\end{figure}

The impact of receiver hardware distortion is investigated in Fig.~\ref{fig:SEvsKappar} for $M = 64$ transmit antennas with $\kappa_{\text{t}} = 0.01$. The distortion-aware precoding is more robust to hardware distortion compared to the power allocation method, with approximately 5 bps/Hz degradation versus approximately 10 bps/Hz for power allocation. Furthermore, transmit distortion has a more significant impact on SE than receive distortion since it directly affects the quality of received symbols.

\newpage

\section{Conclusion} \label{sec:concls}

This paper addressed the operation of ISAC systems under practical hardware impairments and environmental clutter. We derived a closed-form expression for the sensing SCNR that accounts for both transmit and receive hardware distortions.
The derived metric was reformulated to facilitate precoder design. A novel distortion-aware precoding framework was proposed to meet sensing requirements with minimal impact on communication performance. The resulting optimization problem was extended to a power allocation strategy with dedicated sensing signals. Simulation results confirmed that the proposed method maintains the target SCNR, while communication performance remains close to that of ideal hardware schemes. The analysis also revealed how system parameters, such as the number of BS antennas and hardware impairment levels, influence performance. Future work will extend this framework to wideband systems and multi-target scenarios.

\bibliographystyle{IEEEtran}
\bibliography{references}

\end{document}